\documentclass[aps,showpacs,notitlepage,superscriptaddress,longbibliography]{revtex4-1}

\usepackage{epsfig,graphics,amssymb,amsmath,subeqnarray,color,bm,bbm}
\usepackage[toc,page]{appendix}
\usepackage{mathtools}
\usepackage{xcolor}
\usepackage{float}
\usepackage{graphicx}
\usepackage[colorlinks=true,allcolors=blue]{hyperref}
\UseRawInputEncoding
\begin{document}

\title{Effects of direction reversals on patterns of active filaments}
\author{Leila Abbaspour}
\email{leila.abbaspour@mtl.maxplanckschools.de}
\affiliation{Max Planck School Matter to Life, University of G{\"o}ttingen, Friedrich-Hund-Platz 1, 37077 G{\"o}ttingen, Germany}

\affiliation{Institute for Dynamics of Complex Systems, University of G{\"o}ttingen, Friedrich-Hund-Platz 1, 37077 G{\"o}ttingen, Germany}

\author{Ali Malek}

\affiliation{Institute for Dynamics of Complex Systems, University of G{\"o}ttingen, Friedrich-Hund-Platz 1, 37077 G{\"o}ttingen, Germany}

\author{Stefan Karpitschka}
\affiliation{Max Planck Institute for Dynamics and Self-Organization, Am Faßberg 17, 37077 G{\"o}ttingen, Germany}

\author{Stefan Klumpp}
\email{stefan.klumpp@phys.uni-goettingen.de}
\affiliation{Institute for Dynamics of Complex Systems, University of G{\"o}ttingen, Friedrich-Hund-Platz 1, 37077 G{\"o}ttingen, Germany}
\affiliation{Max Planck School Matter to Life, University of G{\"o}ttingen, Friedrich-Hund-Platz 1, 37077 G{\"o}ttingen, Germany}

\date{\today}

%%%%%%%%%%%%%%%%%%%%%%%%%%%%%%%%%%%%
%%%%%%%%%%%%%%%%%%%%%%%%%%%%%%%%%%%%
%%%%%%%%%%%%%%%%%%%%%%%%%%%%%%%%%%%%
%%%%%%%%%%%%%%%%%%%%%%%%%%%%%%%%%%%%
%%%%%%%%%%%%%%%%%%%%%%%%%%%%%%%%%%%%%%%%%%%%%%%%%%%%%%%%%%%%%%%%%%%%%%
%                                                                    %
%                           abstract                                 %
%                                                                    %    
%%%%%%%%%%%%%%%%%%%%%%%%%%%%%%%%%%%%%%%%%%%%%%%%%%%%%%%%%%%%%%%%%%%%%%

\begin{abstract}

Active matter systems provide fascinating examples of pattern formation and collective motility without counterpart in equilibrium systems. Here, we employ Brownian dynamics simulations to study the collective motion and self-organization in systems of self-propelled semiflexible filaments, inspired by the gliding motility of \textit{filamentous Cyanobacteria}. Specifically, we investigate the influence of stochastic direction reversals on the patterns.  We explore pattern formation and dynamics by modulating three relevant physical parameters, the bending stiffness, the activity, and the reversal rate. In the absence of reversals, our results show rich dynamical behavior including spiral formation and collective motion of aligned clusters of various sizes, depending on the bending stiffness and self-propulsion force. The presence of reversals diminishes spiral formation and reduces the sizes of clusters or suppresses clustering entirely. This homogenizing effect of direction reversals can be understood as reversals providing an additional mechanism to either unwind spirals or to resolve clusters. 
\end{abstract}

\maketitle
%%%%%%%%%%%%%%%%%%%%%%%%%%%%%%%%%%%%
%%%%%%%%%%%%%%%%%%%%%%%%%%%%%%%%%%%%
%%%%%%%%%%%%%%%%%%%%%%%%%%%%%%%%%%%%
%%%%%%%%%%%%%%%%%%%%%%%%%%%%%%%%%%%%
%\tableofcontents
%%%%%%%%%%%%%%%%%%%%%%%%%%%%%%%%%%%%%%%%%%%%%%%%%%%%%%%%%%%%%%%%%%%%%%
%                                                                    %
%                           Introduction                             %
%                                                                    %    
%%%%%%%%%%%%%%%%%%%%%%%%%%%%%%%%%%%%%%%%%%%%%%%%%%%%%%%%%%%%%%%%%%%%%%
\section{Introduction}

Active matter is a class of systems that are inherently out of equilibrium through coupling to an internal process of energy consumption that results, for example in self-propulsion or growth \cite{RMP2013,RMP2016}. Specifically, the case of self-propelled active particles has received much attention in recent years, as it gives rise to intriguing dynamics and mechanical behaviour at the collective level. Such systems are ubiquitous in nature, examples include the cytoskeleton~\cite{hyman1996,ndlec1997,surrey2001}, swarming bacteria~\cite{Mendelson1999,parrish2002,dombrowski2004}, tissues~\cite{Salbreux2017} ,and biofilms~\cite{Hall2004}. Finding suitable tools to study active matter systems is a challenge for non-equilibrium statistical physics in order to elucidate the dynamical behavior and the physical properties of these systems. 

Self-propelled particles self-organize into macroscopic structures with collective dynamics like clustering, swarming, and swirling. This type of dynamical pattern formation has been subject to intensive research in recent years, both from a theoretical or computational viewpoint ~\cite{jeffrey2020,joshi2019,review2019,Prathyusha2018,duman2018,shi2018,wang2018,grossman2016,Weitz2015,Balagam2015,Rolf2015,Abkenar2013,Wensink2012,Wensink2008,Leila2021} and experimentally~\cite{Uwe2002,sumino2012,Peruani2012,Rabani2013,Dunkel2013,Avraham2014,Nishiguchi2017,huber2018}, often using systems of bacteria either swimming in a solution or gliding on surfaces as well as synthetic self-propelled particles systems~\cite{Polyarti2007}. 

On the theoretical side, most studies of collective effects in systems of self-propelled particles have considered either spherical or point-like particles   ~\cite{Leila2021}or short and rigid rod-like particles, as these are good approximations for the shape of the bacteria typically used in experiments such as swimming \textit{Bacillus subtilis} or \textit{Escherichia coli} \cite{Nishiguchi2017, review2019,Wensink2012} and as \textit{Myxococcus xanthus}, gliding on surfaces \cite{Peruani2012}. The aspect ratio of rod-like particles has a strong impact on the patterns, as high aspect ratios promote (nematic) alignment of the particles and swarming \cite{Wensink-PNAS-2012}. 

At very large aspect ratios, the the finite rigidity of the particles is expected to play a role. Therefore several computational recent studies have addressed semi-flexible filamentous particles or self-propelled polymers \cite{Prathyusha2018,duman2018}. Indeed, their collective behavior is modulated by their bending rigidity. Moreover, flexibility even results in new single-filament behavior, as individual flexible filaments are seen to curl into spirals due to their self-propulsion \cite{Rolf2015}.

Complicating the understanding of the behavior of the collective motion of microbes, many of them in addition to active self-propulsion also perform active directional changes such as  the well-known run-and-tumble motion of \textit{Escherichia coli}~\cite{Linda2000}. Typically, these active direction changes are an integral part of the mechanisms for chemotaxis and other types of tactic behaviors, through a coupling of the rates of directional change to the direction of motion relative to a (chemical, light, etc.) gradient direction. Such active directional changes have occasionally been included in models for the motility and collective behavior of self-propelled particles \cite{grossman2016,Grosmann2016-2,santra2021,zhang2018} or spreading of active polymers in porous media \cite{Kurzthaler2021}. For  bacteria, the simplest direction change is a reversal, where the whole filament reverses its direction of motion by  $180^{\circ}$, as seen in the gliding motility of \textit{M. xanthus}, a short rigid rod \cite{Peruani2012}, as well as in gliding  filamentous \textit{Cyanobacteria} ~\cite{McBride2001,tam2011}.

In this article, we use agent-based simulations to address how such reversals affect the collective behaviors of self-propelled semi-flexible filaments at overall relatively low density. To that end, we first consider the reference case without reversals and analyze the formation of spirals and swarms or clusters.
Then we turn to the case with stochastic direction reversals, generated by a Poisson process, and investigate their impact on these patterns.

With that in mind, in this paper, we initially examine the self-assembly of active filaments by 
modulating the bending stiffness and self-propulsion force of the filaments, and then investigate how the introduction of reversals modifies the collective behavior.

We focus on the interplay of the filaments'  self-propulsion force, bending stiffness, and reversal rate as the control parameters that modulate the pattern formation process. In the absence of reversals, the filaments form spirals for high self-propulsion force and low bending stiffness, for larger bending stiffness coherently moving clusters dominate the system. Introducing reversals reduces the rates of forming either type of pattern, and thus the system becomes gradually more isotropic with increasing reversal rate. 
The suppression of either pattern can be explained by the dynamical pathway of these structures: First, reversals add a mechanism that interrupts the spooling of a filament onto itself and initiates unwinding. Second, similarly, reversals admit filaments to leave a cluster as they break the coherence of motion in clusters.

%%%%%%%%%%%%%%%%%%%%%%%%%%%%%%%%%%%%%%%%%%%%%%%%%%%%%%%%%%%%%%%%%%%%%%
%                                                                    %
%                           Model                                    %
%                                                                    %    
%%%%%%%%%%%%%%%%%%%%%%%%%%%%%%%%%%%%%%%%%%%%%%%%%%%%%%%%%%%%%%%%%%%%%%

%%%%%%%%%%%%%%%%%%%%%%%%%%%%%%%%%%%%%%%%%%%%%%%%%%%%%%%%%%%%%%%%%%%%%%%%%%%
\begin{figure}[t!]
    \centering
   \includegraphics[width=0.8\linewidth]{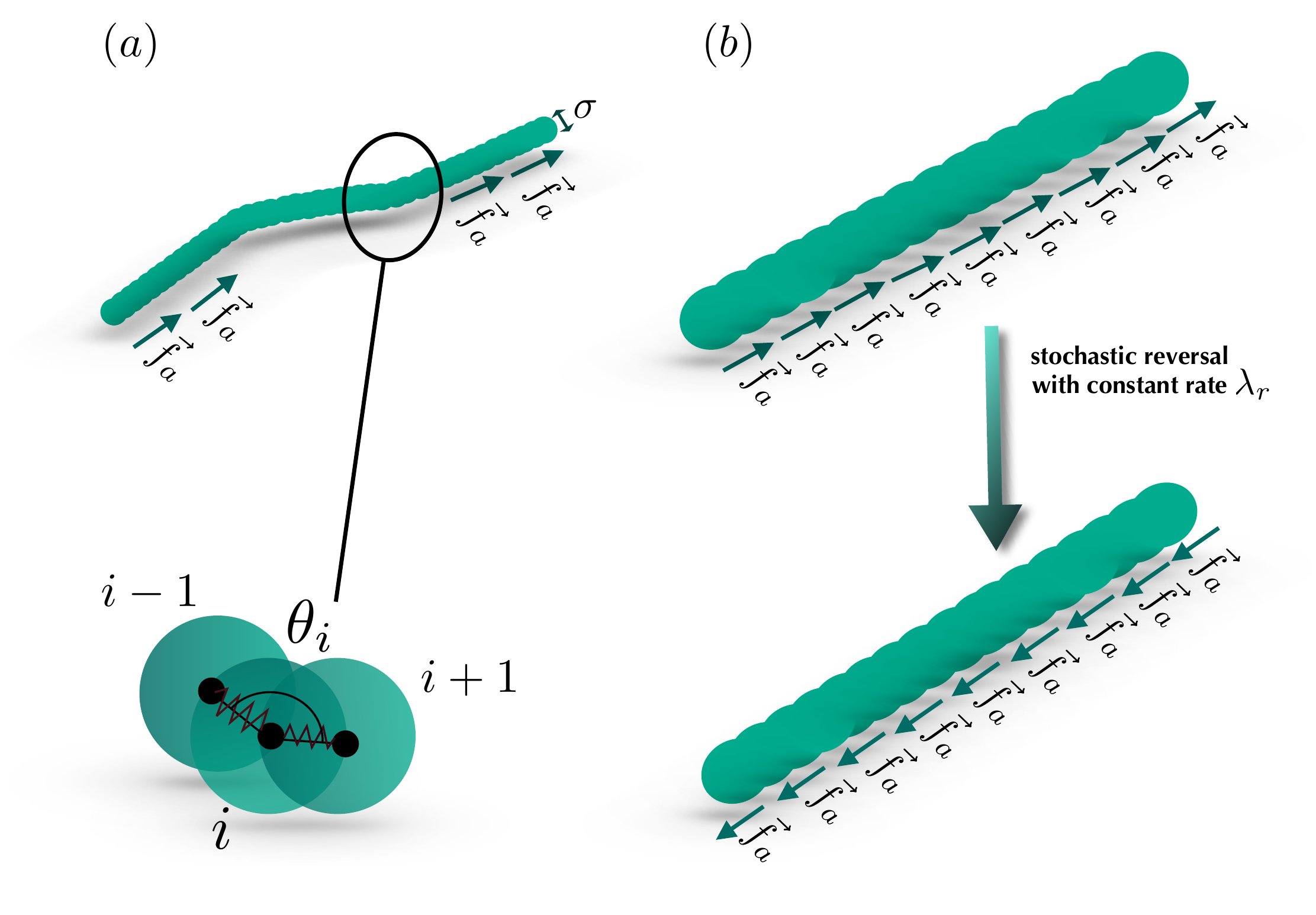}
    \caption{ \textbf{Model of a self-propelled filament. (a)} Sketch of a self-propelled filament composed of monomers with diameter $\sigma$. The active force inducing their self-propulsion  acts tangentially along each bond. \textbf{(Zoom)} The flexibility of the chain is modeled by a harmonic bending potential, where $\theta_i$ is the angle formed by each successive triplet of monomers $i-1,i,i+ 1$. 
    \textbf{(b)} Sketch of the reversal mechanism of a self-propelled filament: Upon a stochastic reversal with rate $\lambda$, a filament reverses the direction of its self-propulsion force by $180^{\circ}$.}
    \label{fig:Model}
\end{figure}
\section{\label{sec:level1}Model}
%%%%%%%%%%%%%%%%%%%%%%%%%%%%%%%%%%%%%%%%%%%%%%%%%%%%%%%%%%%%%%%%%%%%%%%%%%%

\subsection{Self-propelled filaments}

We consider a system of $N_c$ self-propelled semi-flexible filaments. 
The individual filament is described as a chain of $N$ monomers located at position s $(\vec{r}_1, \vec{r}_2, \cdots, \vec{r}_N)$ in space (see Fig.~\ref{fig:Model} (a) for a schematic). Two consecutive  monomers in the chain are connected by a harmonic spring. In addition, the angle between three consecutive monomers along the chain are subject to a harmonic bending potential, which controls the flexibility of the chain. In the over-damped regime, the equation of motion for each monomer can be written as:
\begin{equation}
\zeta\dot{\vec{r}}_i = -\nabla_i U_s - \nabla_i U_b+ \vec{F}^{a}_i + \vec{F}^{ex}_i.
\label{eq:equation of motion}
\end{equation}
where $\dot{\vec{r}}_i$ is the velocity of the $i$-th monomer, $\zeta$ is the friction coefficient. The four force contributions on the right hand side of this equation represent stretching and bending of the chain, the active force responsible for self-propulsion, and volume exclusion, respectively. 

$U_s$ is the stretching potential due to the harmonic springs (each with an equilibrium length of $r_0$ between consecutive monomers) with the form
%%%%%%%%%%%%%%%%%%%%%%%%%%%%%%%%%%%%%%%%%%%%%%%%%%%%%%%%%%%%
\begin{equation}
U_s = \frac{\kappa_s}{2}\sum_{j = 2}^{N} (r_{j,j-1} - r_0)^2.
\end{equation}
%%%%%%%%%%%%%%%%%%%%%%%%%%%%%%%%%%%%%%%%%%%%%%%%%%%%%%%%%%%%
where $r_{j,j-1} = |\vec{r}_j - \vec{r}_{j-1}|$ is the distance between the $j$-th and $j-1$-th monomer and $\kappa_s$ is the spring constant associated with stretching. 

$U_b$ is the harmonic bending potential which controls the flexibility of the bonds. Considering $\theta_i$ as the angle formed by a consecutive triplet of monomers $(i-1, i, i+1)$, given by  $\theta_i=cos^{-1}\left(\frac{{\vec{r}}_{i-1,i} \cdot {\vec{r}}_{i,i+1}}{|{\vec{r}}_{i-1,i}||{\vec{r}}_{i,i+1}|}\right)$, this potential is defined as 
%%%%%%%%%%%%%%%%%%%%%%%%%%%%%%%%%%%%%%%%%%%%%%%%%%%%%%%%%%%%
\begin{equation}
U_b = \frac{\kappa_b}{2}\sum_{j=2}^{N-1} (\theta_j-\pi)^2.
\end{equation}
%%%%%%%%%%%%%%%%%%%%%%%%%%%%%%%%%%%%%%%%%%%%%%%%%%%%%%%%%%%%
where $\pi$ is the equilibrium angle of between adjacent pairs of monomers, corresponding to a straight chain, and $\kappa_b$ is the spring constant associated with bending. 

The self-propulsion is induced by the active force $\vec{F}^{a}_i$,
%%%%%%%%%%%%%%%%%%%%%%%%%%%%%%%%%%%%%%%%%%%%%%%%%%%%%%%%%%%%%%%%%%%%%%%%%%%%%%%%%%%%%%%%%%%%%%%%%%%%%%%%%%%%%%%%%%%%%%%%
\begin{equation}
\vec{F}^{a}_i = f_a \hat{t}_{i},
\label{eq:active force}
\end{equation}
%%%%%%%%%%%%%%%%%%%%%%%%%%%%%%%%%%%%%%%%%%%%%%%%%%%%%%%%%%%%%%%%%%%%%%%%%%%%%%%%%%%%%%%%%%%%%%%%%%%%%%%%%%%%%%%%%%%%%%%%
where $f_a$ fixes the absolute value of the self-propulsion force and $\hat{t}_i$, the unit vector tangent to the chain at the position of $i$-th monomer, defines its direction. For a chain consisting of discrete monomers, this vector can be approximated as:
%%%%%%%%%%%%%%%%%%%%%%%%%%%%%%%%%%%%%%%%%%%%%%%%%%%%%%%%%%%%%%%%%%%%%%%%%%%%%%%%%%%%%%%%%%%%%%%%%%%%%%%%%%%%%%%%%%%%%%%%
\begin{equation}
\hat{t}_i = \frac{\alpha_c}{2}\left(\frac{\vec{r}_{i+1,i}}{|\vec{r}_{i+1,i}|} + \frac{\vec{r}_{i,i-1}}{|\vec{r}_{i,i-1}|}\right).
\label{eq:unit tangent vector}
\end{equation}
%%%%%%%%%%%%%%%%%%%%%%%%%%%%%%%%%%%%%%%%%%%%%%%%%%%%%%%%%%%%%%%%%%%%%%%%%%%%%%%%%%%%%%%%%%%%%%%%%%%%%%%%%%%%%%%%%%%%%%%%
Here $\alpha_c \in \{-1, +1\}$ specifies the state of the polarity (head-tail) for each chain $c \in \{1, \cdots, N_c\}$ where $N_c$ is the number of chains in the system. This variable will be important for the modelling of reversal events, as we will discuss below. 

Finally the presence of the repulsive force $\vec{F}^{ex}_i$ prevents overlaps of the monomers. Such volume exclusion is implemented  using the  Week-Chandler-Anderson potential:
%%%%%%%%%%%%%%%%%%%%%%%%%%%%%%%%%%%%%%%%%%%%%%%%%%%%%%%%%%%%%%%%%%%%%%%%%%%%%%%%%%%%%%%%%%%%%%%%%%%%%%%%%%%%%%%%%%%%%%%%
\begin{equation}
    U_{ext}(r)= 
\begin{cases}
     4 \epsilon \left[ \left(\frac{\sigma}{r}\right)^{12} + \left(\frac{\sigma}{r}\right)^6 \right] + \epsilon,& r < 2^{1/6}\sigma\\
    0,              & \text{otherwise}
\end{cases}
\end{equation}
%%%%%%%%%%%%%%%%%%%%%%%%%%%%%%%%%%%%%%%%%%%%%%%%%%%%%%%%%%%%%%%%%%%%%%%%%%%%%%%%%%%%%%%%%%%%%%%%%%%%%%%%%%%%%%%%%%%%%%%%
where $\sigma$ is the nominal interaction diameter and $\epsilon$ is the energy scale of the interaction. In addition, adjacent monomers in the filament overlapped, resulting in a smooth filament and $r_0=\sigma/2$

\subsection{Direction reversals}

So far, our model agrees with previous models for self-propelled filaments \cite{duman2018,Prathyusha2018,Rolf2015}. In addition, we include spontaneous reversals of the direction of self-propulsion, as they are often seen in the surface motion of bacteria \cite{taktikos2013,Kurzthaler2021} including filamentous species such as gliding filamentous \textit{Cyanobacteria} \cite{hader1982,hader1987,tamulonis2011}. Typically, the rates for such reversals are modulated by various tactic behaviors such as chemotaxis, phototaxis, etc. Here we consider the baseline case of spontaneous reversals in a homogeneous environment.

We model random reversal events, where the gliding direction of a chain changes instantaneously by $180^{\circ}$, as generated from a Poisson process with  a reversal rate $\lambda_r$ (see Fig.~\ref{fig:Model}b). Mathematically, this means that the polarity (head-tail) state $\alpha_c$ of a chain $c$ follows a stochastic process
\begin{equation}
\alpha_c=1\quad \xrightleftharpoons[\lambda_r]{\lambda_r}\quad \alpha_c=-1.
\end{equation}
In such a Poisson process, the distribution of waiting times $\tau$ elapsed between two consecutive reversals is an exponential distribution of the form
%%%%%%%%%%%%%%%%%%%%%%%%%%%%%%%%%%%%%%%%%%%
\begin{equation}
P_{\rm w}(\tau) =\lambda_r e^{-\lambda_r \tau}.
\label{eq:reversal_process}
\end{equation}
%%%%%%%%%%%%%%%%%%%%%%%%%%%%%%%%%%%%%%%%%%%
Hence to implement the reversals for a single chain, we draw a random waiting time $\tau_0$ from the exponential distribution described in Eq. \ref{eq:reversal_process} at time $t_0$ . Between $t_0$ and $t_1 = t_0 + \tau_0$, the chain evolves following Eq. \ref{eq:equation of motion}. At $t_1$ the gliding direction of the chain is reversed and another waiting time $\tau_1$ is drawn from the same waiting time distribution. This process is done for each chain independently during the simulation by drawing subsequent time intervals $\tau_2,\tau_3,...$, such that reversals of any chain in the system is uncorrelated with the reversal of any other chains.

Given a reversal rate $\lambda_r$ and a free propulsion speed of $f_a/\zeta$, the free propagation distance between reversals on average is $L_r = f_a/(\zeta\,\lambda_r)$.

\subsection{Simulations and parameters}

All simulations presented here were run on GPUs using the simulation package HOOMD-blue \cite{HMD} with custom extensions for self-propulsion and for reversals, which were compiled together with the whole software. The simulation integrates Eq. \ref{eq:equation of motion} over $t=10^8$ time steps.

In these simulations, lengths are measured in units of the monomer radius $\sigma$ and the energy unit is $k_BT$. We choose the time unit to be the self-diffusion time for a single monomer $\tau_D =\frac{\sigma^2 \zeta}{k_BT}$ where $\zeta$, which determines the damping, is set to $\zeta=15$. Our system consists of $N_f=1666$ filaments each having $N_m=59$ monomers.

We set the spring constant of the stretching spring to $\kappa_s=10^4$, {\textit{i.e.}} to a relatively high value to make sure that chains do not stretch too much along their axis. Hence the distance between two successive monomers does not fluctuate strongly and is approximately  constrained to the fixed value $r_0$. Furthermore, to obtain a smoother filament, we also set the value $r_0=\frac{\sigma}{2}$. To investigate a large parameter space, we varied $\kappa_b$ in the range $\{1, 1800\}$ and $f_a$ between $\{1, 100\}$. The packing fraction, $\phi=N_f N_m r_0 \sigma/L^2$, is set to $0.1$ in all simulations, where $L$ denotes the box size and $L=167\sigma$ 

%and N denotes the total number of monomers in the system, $N=N_f.N_m$.
 
The aspect ratio, a dimensionless number representing the ratio between the contour length and the diameter of the individual monomers  $a=\frac{L_f}{\sigma}$, is set to almost $30$ and $L_f=(N_m+1) r_0$ is the contour length of the filaments.  The reversal rates $\lambda_r$ are varied between $\{0-0.1\}$ and are same for all the filaments.

%%%%%%%%%%%%%%%%%%%%%%%%%%%%%%%%%%%%%%%%%%%%%%%%%%%%%%%%%%%%%%%%%%%%%%
%                                                                    %
%                           Results                                  %
%                                                                    %    
%%%%%%%%%%%%%%%%%%%%%%%%%%%%%%%%%%%%%%%%%%%%%%%%%%%%%%%%%%%%%%%%%%%%%%

\section{Results}

To address the effect of direction reversals on the collective dynamics of self-propelled filament, we performed systematic simulations varying three key parameters, the bending stiffness $\kappa_b$, the strength of self-propulsion $f_a$, and the reversal rate $\lambda_r$. We will show typical snapshots of these simulations below. We start with the case without direction reversals ($\lambda_r=0$), which has been studied before \cite{Prathyusha2018,duman2018} and which serves us as a reference scenario here. In that case, it is known that a variety of non-equilibrium patterns can be formed, for which we will present a detailed diagram of states below.

Later on, we add the reversal mechanism to the filaments, inspired by the reversal dynamics (due to phototaxis) observed in \textit{filamentous Cyanobacteria} to investigate the response of the system to an abrupt and random reversal in gliding direction. We report a dramatic change in the individual as well as collective dynamics in the state diagram as we crank up the reversal rate, eventually destroying the spiral state completely and leading to destruction of clusters and collective motion.
%%%%%%%%%%%%%%%%%%%%%%%%%%%%%%%%%%%%%%%%%%%%%%%%%%%%%%%%%%%%%%%%%%%%%%
%                                                                    %
%                           Non-reversing active filaments           %
%                                                                    %    
%%%%%%%%%%%%%%%%%%%%%%%%%%%%%%%%%%%%%%%%%%%%%%%%%%%%%%%%%%%%%%%%%%%%%% 
\subsection{Non-reversing active filaments}
 
%%%%%%%%%%%%%%%%%%%%%%%%%%%%%%%%%%%%%%%%%%%%%%%%%%%%%%%%%%%%%%%%%%%%%%
%                                                                    %
%                           Spiral formation                         %
%                                                                    %    
%%%%%%%%%%%%%%%%%%%%%%%%%%%%%%%%%%%%%%%%%%%%%%%%%%%%%%%%%%%%%%%%%%%%%%

%%%%%%%%%%%%%%%%%%%%%%%%%%%%%%%%%%%%%%%%%%%%%%%%%%%%%%%%%%%%%%%%%%%%%%%%%%%%%%%%%%%%%%%%%%%%%%%%%%%%%%%%%%%%%%%%%%%%%%%%
\begin{figure}[t]
    \centering
   \includegraphics[width=0.8\linewidth]{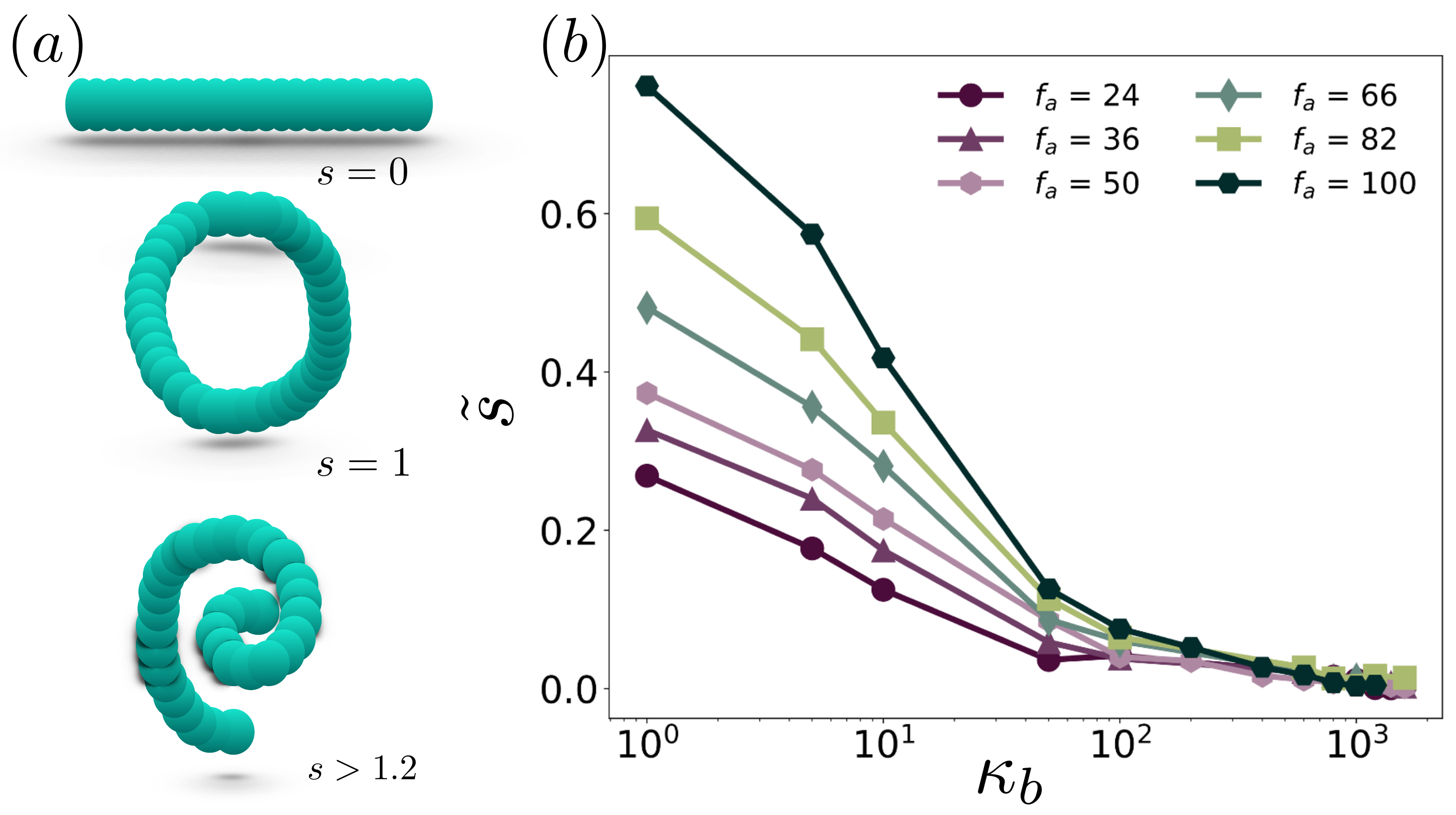}
    \caption{\textbf{Spiral ratio as a function of bending stiffness for different values of self-propulsion force.} \textbf{(a)} Sketch of different configuration of a single filament characterised by the indicated spiral number indicated by $s$. \textbf{(b)} Quantification of spiral formation by the spiral ratio $\tilde{s}$: An increase in bending stiffness decreases the probability of spiral formation at a constant self-propulsion force. Different values for self-propulsion are shown in different colors. With increasing self-propulsion force at low bending stiffness, spiral formation is promoted. For high bending stiffness, spiral formation does not take place regardless of the self-propulsion force. The aspect ratio of the filaments is  $a=30$ and the packing fraction is $\phi = 0.1$.}
    \label{fig:Spiral-ratio}
\end{figure}
%%%%%%%%%%%%%%%%%%%%%%%%%%%%%%%%%%%%%%%%%%%%%%%%%%%%%%%%%%%%%%%%%%%%%%%%%%%%%%%%%%%%%%%%%%%%%%%%%%%%%%%%%%%%%%%%%%%%%%%%%

\subsubsection{Spiral formation}

In the absence of direction reversals, the collective dynamics of the chains is dominated by the interplay between self-propulsion, bending stiffness and confinements due to the excluded volume effect. A key observation at low density of filaments is that flexible filaments may form spirals while stiffer filaments tend to form clusters \cite{Prathyusha2018,duman2018}.  

Spirals form due to the self-interaction of an isolated active filament with high self-propulsion and low bending stiffness. Spiral formation stands in contrast to the relatively straight equilibrium structure of a semi-flexible filament and depends on the self-propulsion. Spiral formation is initiated when the head of the filament collides with a  subsequent part of its own body, such that the excluded volume interaction forces the filament to wind into itself. If the self-propulsion is strong enough, filaments get trapped in their own body, forming stable spirals. To characterise the dynamical state with spiral formation we define a spiral number~\cite{Rolf2015} for each filament in the system which quantifies how many times a filament is wound up around itself and is defined as:
%%%%%%%%%%%%%%%%%%%%%%%%%%%%%%%%%%%%%%%%%%%%%%%%%%%%%%%%%%%%%%%%%%%%%%%
\begin{equation}
s = \left| \frac{\sum_{i=2}^{N-2} \Delta \theta_i}{2\pi} \right|
\end{equation}
%%%%%%%%%%%%%%%%%%%%%%%%%%%%%%%%%%%%%%%%%%%%%%%%%%%%%%%%%%%%%%%%%%%%%%%
where $\theta_i$ is the angle formed by the beads $i-1, i, i+1$. $|..|$ denotes the absolute value of the spiral number since they can be twisted either in the clockwise or counterclockwise direction.  $s = 0$ indicates an almost straight configuration (see Fig.~\ref{fig:Spiral-ratio} (a)) and $s = 1$ a ring shape (see Fig.~\ref{fig:Spiral-ratio} (a)). We consider a filament to be in a  spiral state if $s\geq 1.2$ (see Fig.~\ref{fig:Spiral-ratio}(a)).  

To distinguish global patterns of filaments, we introduce the spiral ratio as the ratio of the number of filaments in the spiral state (i.e., with $s\geq 1.2$), $N_s$, to that in the straight configuration,$N_{sf}$:
%%%%%%%%%%%%%%%%%%%%%%%%%%%%%%%%%%%%%%%%%
\begin{equation}
\tilde{s} = \frac{N_s}{N_{sf}}.
\end{equation}
%%%%%%%%%%%%%%%%%%%%%%%%%%%%%%%%%%%%%%%%%
The spiral ratio is plotted in Fig.\ref{fig:Spiral-ratio}(b) as a function of the bending stiffness and the self-propulsion force. The figure shows a pronounced decrease of the spiral ratio with increasing bending stiffness, which reflects the fact that increasing bending stiffness ($\kappa_b$) hinders the self-interaction needed for spiral formation.  Fig.\ref{fig:Spiral-ratio}(b) also shows that, for fixed bending rigidity, spiral formation is promoted by an increase in the self-propulsion force.

%%%%%%%%%%%%%%%%%%%%%%%%%%%%%%%%%%%%%%%%%%%%%%%%%%%%%%%%%%%%%%%%%%%%%%
%                                                                    %
%                           Cluster formation                        %
%                                                                    %    
%%%%%%%%%%%%%%%%%%%%%%%%%%%%%%%%%%%%%%%%%%%%%%%%%%%%%%%%%%%%%%%%%%%%%%

%%%%%%%%%%%%%%%%%%%%%%%%%%%%%%%%%%%%%%%%%%%%%%%%%%%%%%%%%%%%%%%%%%%%%%%%%%%%%%%%%%%%%%%%%%%%%%%%%%%%%%%%%%%%%%%%%%%%%%%%%%%
\begin{figure}[t]
    \centering
   \includegraphics[width=\linewidth]{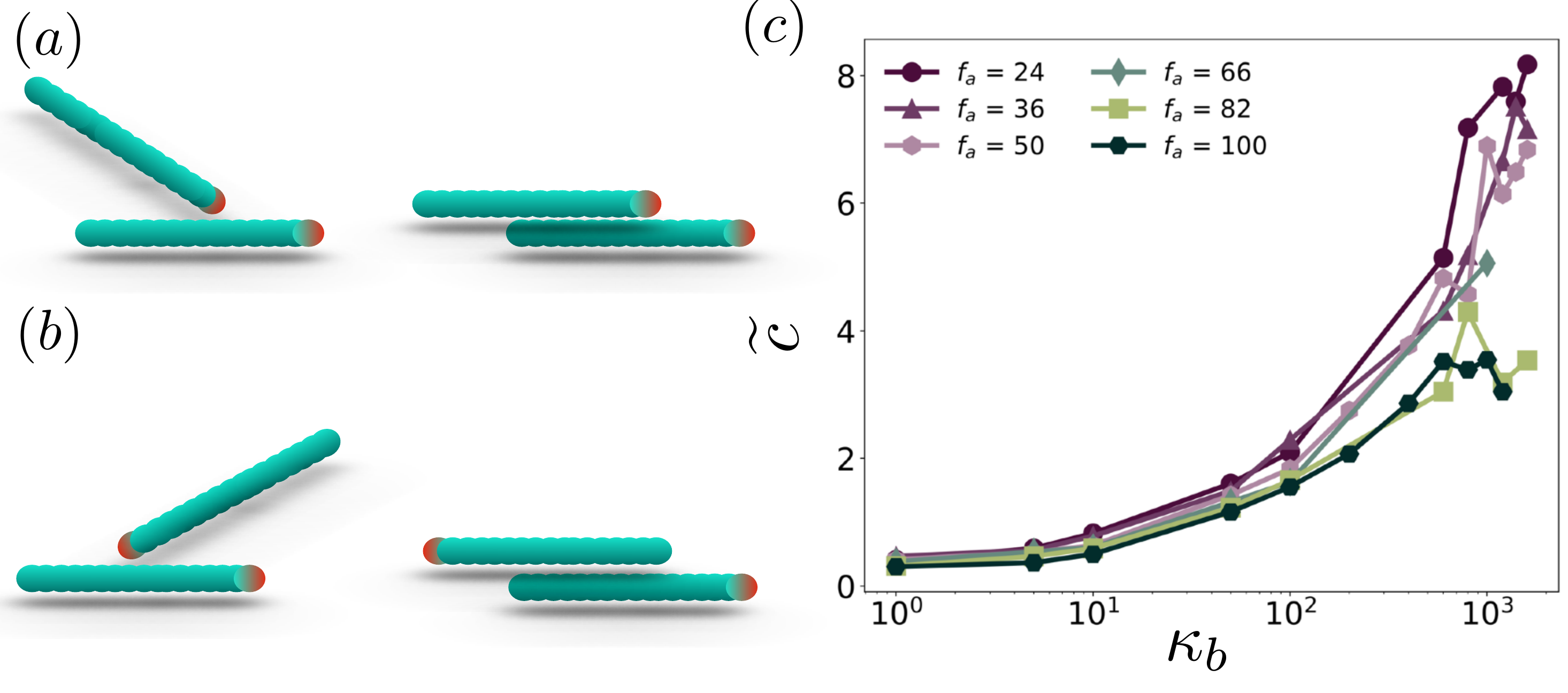}
    \caption{ \textbf{Cluster ratio as a function of bending stiffness for different values of self-propulsion force.}
    \textbf{(a)} A binary collision of self-propelled filaments that  results in parallel alignment and coherent motion that initiates cluster formation. \textbf{(b)} A binary collision of self-propelled filaments that results in anti-parallel alignment and transient coherent motion. The red monomer shows the head of the filament in both cases. \textbf{(c)} Increasing the bending stiffness promotes cluster formation in the system. Different values for the self-propulsion force are shown in different colors. The aspect ratio of the filaments is  $a=30$ and the packing fraction  $\phi = 0.1$.}
    \label{fig:Cluster-ratio}
\end{figure}
%%%%%%%%%%%%%%%%%%%%%%%%%%%%%%%%%%%%%%%%%%%%%%%%%%%%%%%%%%%%%%%%%%%%%%%%%%%%%%%%%%%%%%%%%%%%%%%%%%%%%%%%%%%%%%%%%%%%%%%%%%%
\subsubsection{Cluster formation}

In the limit of high bending stiffness, filaments behave like rod-shaped agents. In this regime, filaments move in the direction of their long axis and, upon collisions, they rotate to align due to their steric interaction. At high bending stiffness, collisions between filaments are dominant over self-interactions as the bending energy required for the self-interaction is rarely reached. Dependent on the angle of incidence of the collision event, the two filaments align either parallel   or anti-parallel (Fig.~\ref{fig:Cluster-ratio}(a) and (b), respectively). In the first case, they continue their motion together, in the second case, they stay in contact transiently  \cite{RMP2013,grossman2016}. 
Clusters of filaments form when additional filaments collide with already aligned filaments, resulting in groups of filaments that  move coherently. The size and life-time of these clusters depends on bending rigidity and activity, as well as the density of the filaments in the system ~\cite{Wensink2012,Wensink-PNAS-2012,RMP2013}.

To characterise the clustering behaviour we identify clusters using the following criteria: two monomers are considered to be part of the same cluster if their centers are distanced less than $1.2\sigma$. We note that this criterium automatically includes all monomers of one filament in the same cluster. In addition, we only considered clusters with at least $N_f=10$ filaments to avoid counting micro clusters. Clusters are identified with the data analysis framework Freud \cite{freud}. Finally, in analogy to the spiral ratio, we define a cluster ratio as the ratio of number of filaments in clusters, $N_c$ to the number of free filaments, $N_f$. 
%%%%%%%%%%%%%%%%%%%%%%%%%%%%%%%%%%%%%%%%%
\begin{equation}
\tilde{c} = \frac{N_c}{N_f}
\end{equation}
%%%%%%%%%%%%%%%%%%%%%%%%%%%%%%%%%%%%%%%%%
\begin{figure}[tb!]
    \centering
   \includegraphics[width=0.5\linewidth]{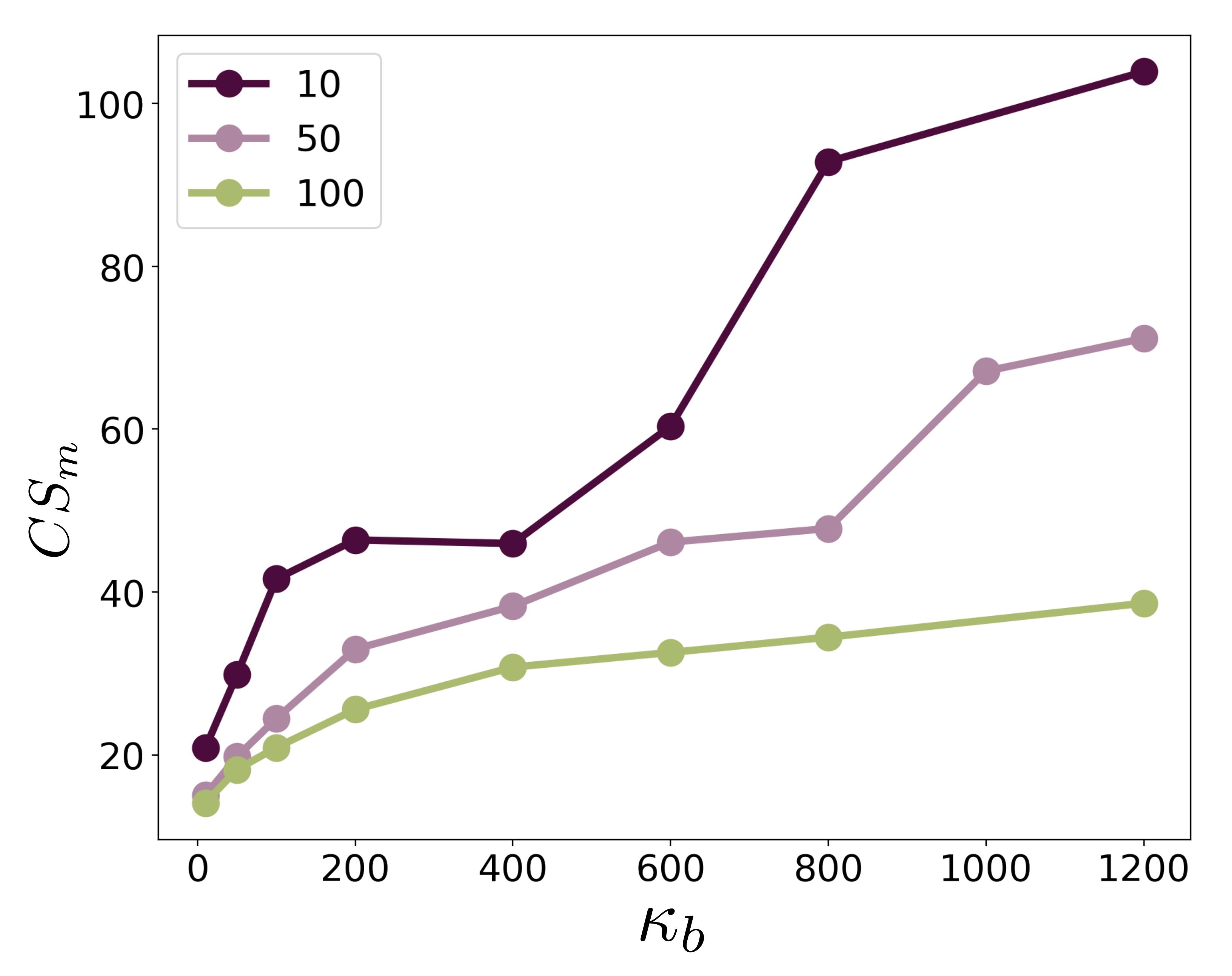}
    \caption{\textbf{Average cluster size as a function of bending stiffness}. The average cluster size is shown in different colors for different self-propulsion forces. }
    \label{fig:mean-cluster-size}
\end{figure}

Fig.~\ref{fig:Cluster-ratio} shows the cluster ratio as a function of the bending stiffness for different values of the  active force. The cluster ratio is seen to increase monotonically  with increasing bending stiffness ($\kappa_b$) and also with increasing self-propulsion force. One can also see that high values of the spiral number are associated with small clusters and vice versa: the regions with higher values of spiral ratio in Fig.~\ref{fig:Spiral-ratio} correspond to the region of low cluster ratio in Fig.~\ref{fig:Cluster-ratio}. When the bending stiffness is low, spirals are formed due to the self-interaction, when the bending stiffness increases, the deformation of the filaments becomes unfavorable, which is why they usually have a straight, rod-like conformation that  promotes alignment and, thus, clustering.

We also define the average cluster size, $CS_m$, which is the average total number of filaments belonging to clusters and this seems to increase with increasing bending stiffness indicating (see Fig.~\ref{fig:mean-cluster-size}) the fact that bending stiffness which suppress spiral formation, promotes clustering.

%%%%%%%%%%%%%%%%%%%%%%%%%%%%%%%%%%%%%%%%%%%%%%%%%%%%%%%%%%%%%%%%%%%%%%
%                                                                    %
%                           Flexture number                          %
%                                                                    %    
%%%%%%%%%%%%%%%%%%%%%%%%%%%%%%%%%%%%%%%%%%%%%%%%%%%%%%%%%%%%%%%%%%%%%%
%%%%%%%%%%%%%%%%%%%%%%%%%%%%%%%%%%%%%%%%%%%%%%%%%%%%%%%%%%%%%%%%%%%%%%%%%%%%%%%%%%%%%%%%%%%%%%%%%%%%%%%%%%%%%%%%%%%%%%%%%%%

Since both spiral ratio and cluster ratio show opposing trends as function of the bending rigidity and of the self-propulsion force, we wondered whether the results could be written as functions of a dimensionless parameter combination. A natural candidate is the flexure number, the ratio of activity and bending rigidity \cite{Sekimoto1995,Winkler2017,duman2018,Prathyusha2018} can be written as:
\begin{equation}
    \Im=\frac{f_aL^3}{\kappa_b}
\end{equation}

\begin{figure}[t!]
    \centering
   \includegraphics[width=\linewidth]{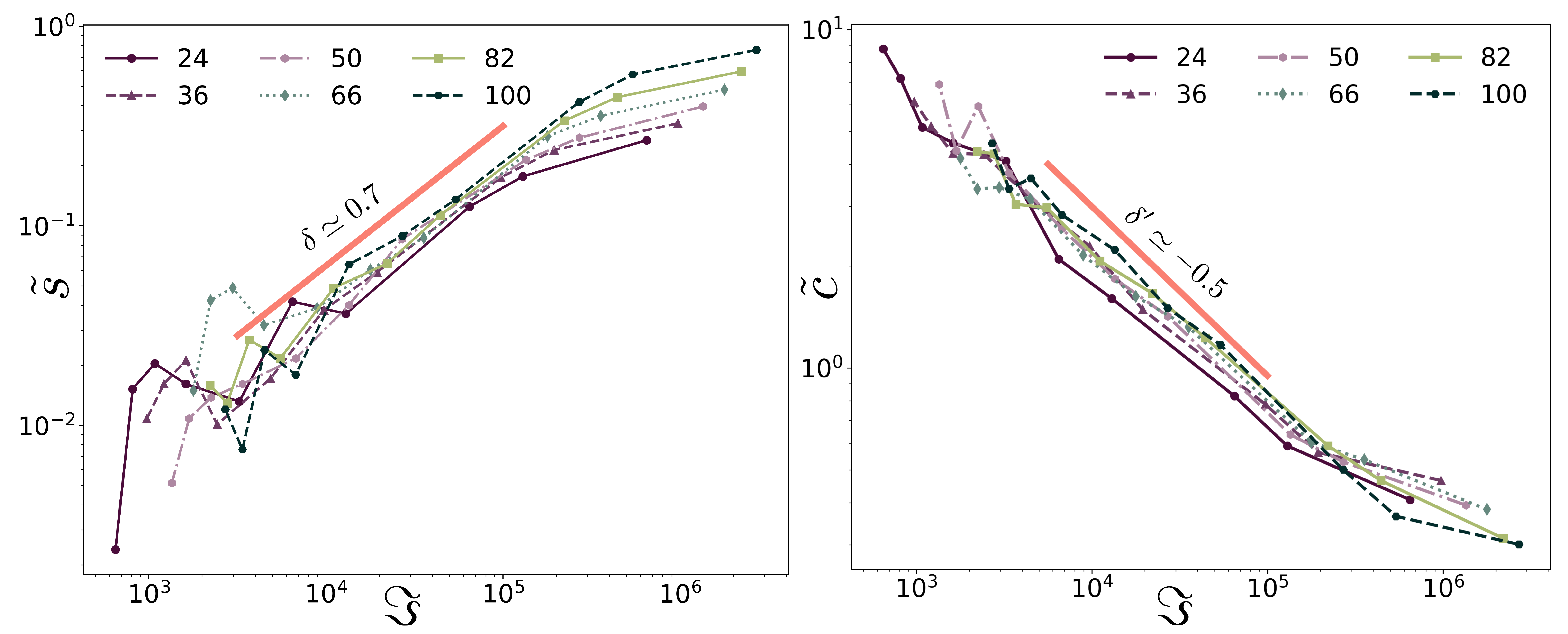}
    \caption{(a) Spiral ratio $\tilde S$ and (b) cluster ratio $\tilde C$ as functions of the flexure number $\Im$: Data collapse for different self-propulsion forces and bending rigidities when these are plotted as functions of a dimensionless combination, the flexure number. The color of the points indicates the value of the self-propulsion force. The red lines indicate power laws  fitted to the curves, with exponents $\delta$ and $\delta'$.}
    \label{fig:flexture}
\end{figure}

Figure \ref{fig:flexture} shows the spiral and cluster ratio as functions of the flexure number. Indeed the data from simulations with different bending stiffness and self.propulsion forces are seen to collapse onto one curve. Only for the largest flexure numbers, we see systematic deviations for the spiral ratio.  For smaller values of the flexure number, the spiral ratio increase as a power law $\tilde S \sim \Im^\delta$ with $\delta\simeq 0.7$. Likewise, the decrease of the cluster ratio also follows a power law $\tilde C \sim \Im ^{\delta'}$ with $\delta'\simeq -0.5$.

%%%%%%%%%%%%%%%%%%%%%%%%%%%%%%%%%%%%%%%%%%%%%%%%%%%%%%%%%%%%%%%%%%%%%%
%                                                                    %
%                           Phase diagram                            %
%                                                                    %    
%%%%%%%%%%%%%%%%%%%%%%%%%%%%%%%%%%%%%%%%%%%%%%%%%%%%%%%%%%%%%%%%%%%%%%
%%%%%%%%%%%%%%%%%%%%%%%%%%%%%%%%%%%%%%%%%%%%%%%%%%%%%%%%%%%%%%%%%%%%%%%%%%%%%%%%%%%%%%%%%%%%%%%%%%%%%%%%%%%%%%%%%%%%%%%%%%%
\begin{figure}[ht!]
    \centering
   \includegraphics[width=0.7\linewidth]{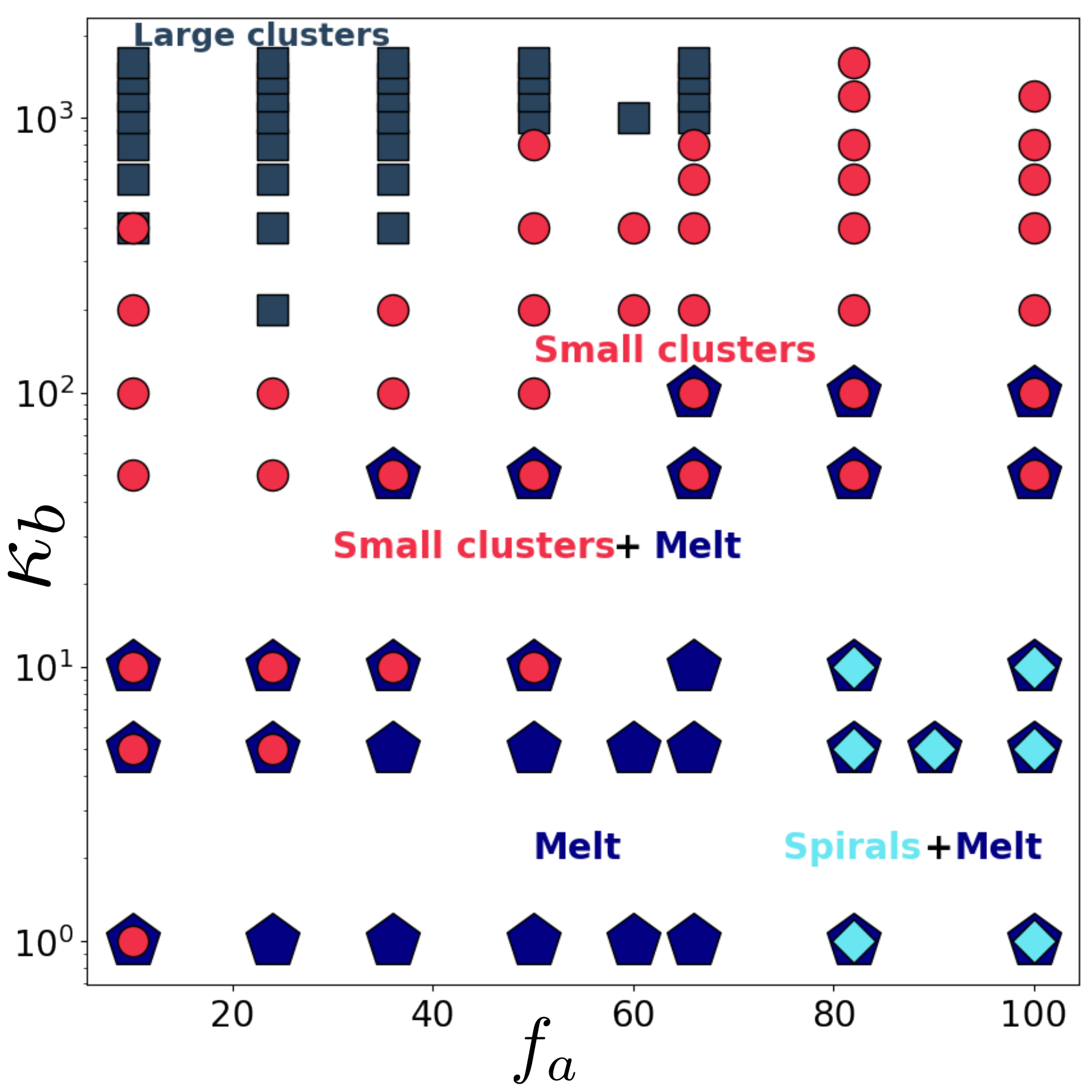}
    \caption{\textbf{Non equilibrium state diagram of non-reversing self-propelled filaments.} The state diagram is drawn as a function of self-propulsion force $f_a$ and the bending stiffness $\kappa_b$ in the absence of reversals. The different states are indicated by different symbols and the color of each symbol is matched to the corresponding designation of the state. For details on the characterisation and identification of the states, see text.} 
    \label{fig:Phase-noreverse}
\end{figure}
%%%%%%%%%%%%%%%%%%%%%%%%%%%%%%%%%%%%%%%%%%%%%%%%%%%%%%%%%%%%%%%%%%%%%%%%%%%%%%%%%%%%%%%%%%%%%%%%%%%%%%%%%%%%%%%%%%%%%%%%%%%
\subsubsection{State diagram}

We performed systematic simulations varying the self-propulsion force $f_a$ and the bending stiffness $\kappa_b$ to  explore the variety of patterns in the steady state of the system. 
 
To distinguish different states of the system, we assign a state to all filaments in a snapshot via the following criteria: A filament is considered as a spiral if its spiral number is $<1.2$. The filament is taken to be in the cluster category if it is not a spiral and part of a cluster of $\geq 10$ filaments. Since this criterion does not capture the diversity of cluster patterns, we further subdivide the cluster state by the size of clusters and distinguish a small-cluster category for filaments in clusters with size $\geq 10$ and $<100$ and a large-cluster category for filaments in clusters with $\geq 100$ filaments. All other filaments are considered to be in the melt state.  

\begin{figure*}[ht!]
    \centering
   \includegraphics[width=0.95\linewidth]{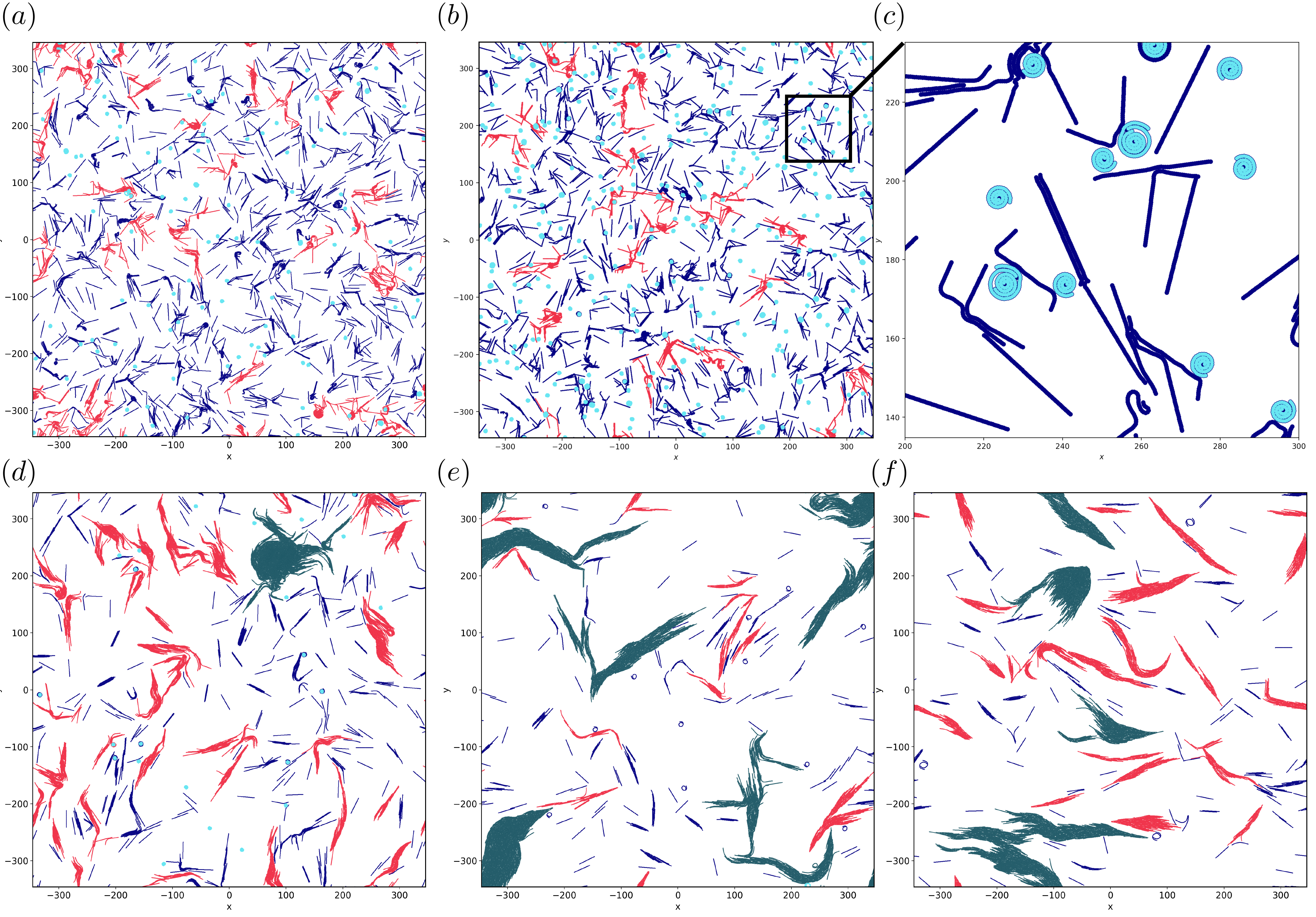}
    \caption{\textbf{Snapshots of the simulation for different $f_a$ and $\kappa_b$}. \textbf{(a)} snapshot for $\kappa_b =10$, $f_a=10$ in `melt' state \textbf{(b)} $\kappa_b =5$, $f_a=100$ in spiral+melt state \textbf{(c)} Close-up view of spirals in plot (a) \textbf{(d)}$\kappa_b =600$, $f_a=82$ \textbf{(e)} in small-cluster state $\kappa_b =1000$, $f_a=24$ \textbf{(f)} in large-cluster state $\kappa_b =1600$, $f_a=10$  in large + small cluster state.}
    \label{fig:Snap-noreverse}
\end{figure*}

Using this classification and averaging over 200 snapshots per condition, we classify the state of the whole system into the following categories:  A large-cluster state or a small-cluster state if $\geq 40\%$ of the filaments are in large or small clusters, respectively. As the two criteria are not mutually exclusive, coexistence of small and large clusters is identified if both are satisfied simultaneously. Likewise, 
a spiral state is identified when more than $20\%$ of the filaments form spirals (the threshold is chosen lower than for the other states, as under the conditions simulated here, we hardly ever see $40\%$ of the filaments being spirals). Finally, the system state is classified as a melt state if $\geq 40\%$ of the filaments fall into this category. Just like for small and large clusters, coexistence of other states is possible and indeed seen frequently, e.g. between melt and small clusters and melt and spirals. 
The diagram of states shown in Fig.~\ref{fig:Phase-noreverse} summarizes this classification. Simulation snapshots corresponding to the different states are shown in Fig.~\ref{fig:Snap-noreverse}, where the color code of the filament shows the classification at the filament level. 

%%%%%%%%%%%%%%%%%%%%%%%%%%%%%%%%%%%%%%%%%%%%%%%%%%%%%%%%%%%%%%%%%%%%%%%%%%%%%%%%%%
%%%%%%%%%%%%%%%%%%%%%%%%%%%%%%%%%%%%%%%%%%%%%%%%%%%%%%%%%%%%%%%%%%%%%%%%%%%%%%%%%%

For low bending stiffness and low self-propulsion force, the filaments are mostly in the melt (or isotropic) phase and move in all directions, as can be seen in Fig.~\ref{fig:Snap-noreverse} (a). We note that while the melt filaments are dominant, there are also some small clusters.  Increasing the self-propulsion force, while the bending stiffness is kept at a small value, promotes spiral formation,  as shown in Fig.~\ref{fig:Snap-noreverse} (b) and (c). Formation of spirals is quite natural in this regime because the filaments are flexible enough to bend onto themselves. Their self-propulsion then results in spiral formation and, once formed, a spiral is typically stable until another filament  collides with it.

Collisions with other filament seems to be the dominant (if not only) mechanism that breaks up spirals, since we have not included explicit translational or rotational noise in our simulations.  Under the conditions we simulated,  the spiral state typically coexists with the melt phase, as can be seen in the snapshots in Fig.~\ref{fig:Snap-noreverse} (b) and (c). Consistent with this observation, a pure spiral state, the "gas of spirals" reported in earlier studies \cite{duman2018} is only observed at even lower densities than used here. 

As shown above (Figure \ref{fig:Spiral-ratio}), the fraction of spirals decreases with increasing bending rigidity. 
For sufficiently large bending stiffnesses, collective behavior via the formation of clusters becomes dominant over individual filament dynamics. In this regime, the self-interaction is less important compared to inter-filament interaction, and when filaments collide, they can bundle into small clusters with local alignment (nematic order). Over time, the clusters grow as other filaments collide with them and join them. At intermediate bending stiffness, there is a coexistence of dispersed individual filaments and small clusters (Fig.~\ref{fig:Snap-noreverse} (d)). As we increase the bending stiffness (Fig.~\ref{fig:Phase-noreverse}d-f), the clusters become larger and the system enters a large-cluster state, in particular for low  self-propulsion force.

These observations show that the collective behavior of filaments is modulated by their bending stiffness and their self-propulsion force. In agreement with earlier work \cite{Prathyusha2018,duman2018}, we distinguish several regimes, characterized by the formation of spirals and of clusters of various sizes. Generally, we observed that, in the absence of direction reversals, filaments with high self-propulsion force and low bending stiffness form spirals, while filaments with high bending stiffness tend to form clusters. This general picture will serve as a reference in the following section, where we will include direction reversals and study how the variation of the reversal rate affects the patterns formed by self-propelled filaments.

%%%%%%%%%%%%%%%%%%%%%%%%%%%%%%%%%%%%%%%%%%%%%%%%%%%%%%%%%%%%%%%%%%%%%%
%                                                                    %
%                         Reversal Rate                              %
%                                                                    %    
%%%%%%%%%%%%%%%%%%%%%%%%%%%%%%%%%%%%%%%%%%%%%%%%%%%%%%%%%%%%%%%%%%%%%%

%%%%%%%%%%%%%%%%%%%%%%%%%%%%%%%%%%%%%%%%%%%%%%%%%%%%%%%%%%%%%%%%%%%%%%%%%%%%%%%%%%%%%%%%%%%%%%%%%%%%%%%%%%%%%%%%%%%%%%%%%%%
\subsection{Active filaments with direction reversals}
\begin{figure*}[htbp]
    \centering
   \includegraphics[width=0.7\linewidth]{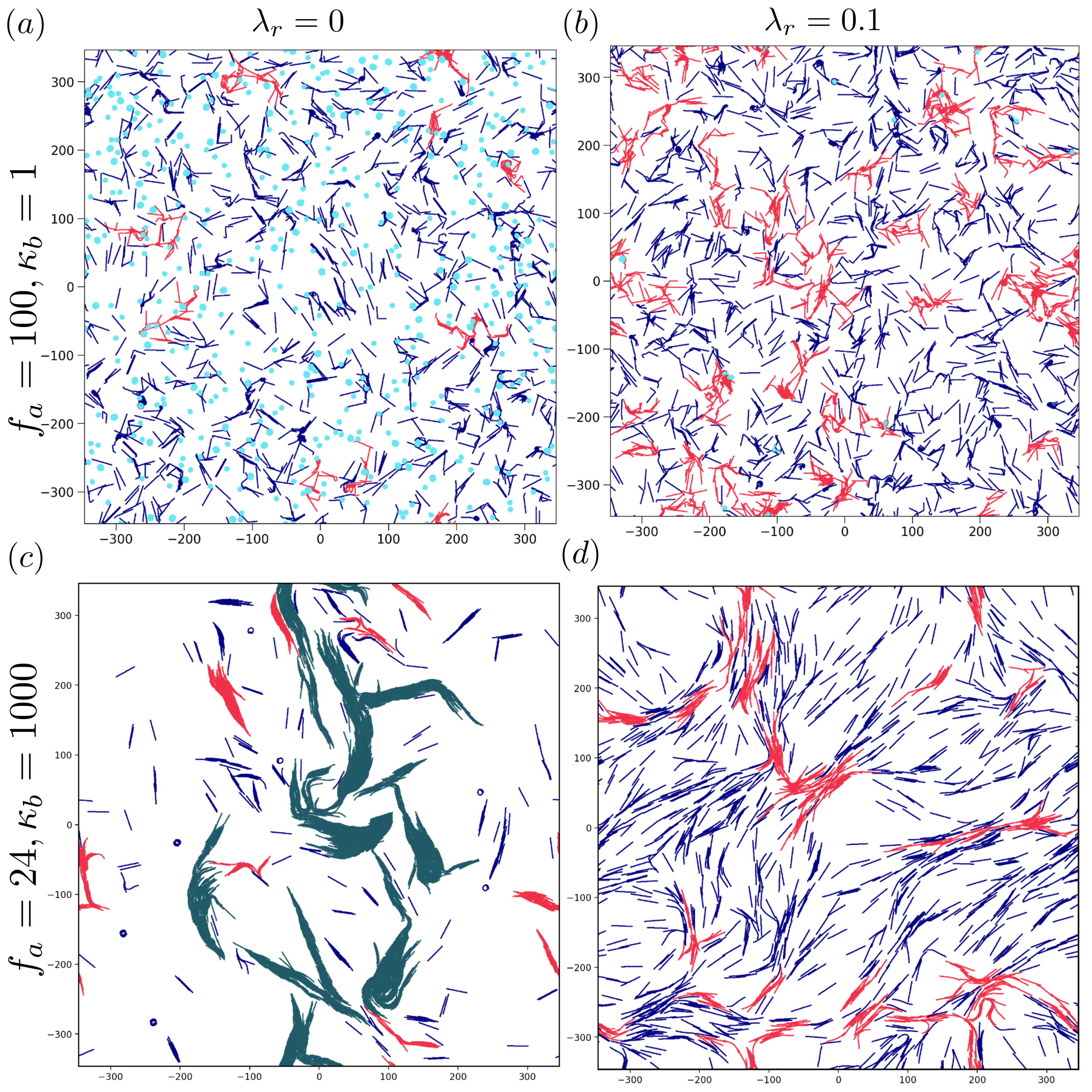}
    \caption{\textbf{Snapshot of the simulation with reversal dynamics.} We show snapshots from the simulations keeping the bending stiffness, activity, density, and aspect ratio same and changing the reversal rate. Applying reversal rate changes the pattern formation drastically from its non-reversing counterpart. \textbf{a,b} in the presence of a nonzero reversal rate, spiral formation has disappeared. \textbf{c,d} Likewise, in the presence of reversals, the cluster sizes become smaller than the non-reversing case. The coloring of the filaments indicates their classification as spirals (cyan), filaments in small clusters (red), large clusters (grey) and a melt (blue). }
    \label{fig:Snap-reverse}
\end{figure*}
%%%%%%%%%%%%%%%%%%%%%%%%%%%%%%%%%%%%%%%%%%%%%%%%%%%%%%%%%%%%%%%%%%%%%%%%%%%%%%%%%%%%%%%%%%%%%%%%%%%%%%%%%%%%%%%%%%%%%%%%%%%
Next, we consider active filaments that exhibit reversals of their direction of motion. Many microorganisms exhibit such a feature in their motility (both swimming and surface-bound species), often as part of their chemo- or photo-tactic strategy. Here we consider the case of spontaneous, non-biased stochastic reversals. Our filaments are taken to change their direction of motion abruptly, by $180^\circ$, according to a Poisson process with rate $\lambda_r$. Varying $\lambda_r$, we explored the effect of reversals on the structures and the collective motion of the active filaments.

Fig.~\ref{fig:Snap-reverse} demonstrates the effect of a small reversal rate $\lambda_r=0.1$ (right column) compared to the case without reversals (left column) for two opposite corners of the state diagram Fig.~\ref{fig:Phase-noreverse}. These two parameters sets chosen here correspond to the regime exhibiting spirals ($f_a=100$ and $\kappa_b=1$, top) and to the regime dominated by small clusters ($f_a=24$ and $\kappa_b=1000$, bottom). In both cases, the visual appearance of patterns clearly changes, indicating a substantial influence of the reversals.
In the spiral+melt regime (large $f_a$ and small $\kappa_b$), spirals (light blue) are abundant (though in co-existence with straight filaments) in absence of reversals, but efficiently suppressed by a small reversal rate $\lambda_r=0.1$.
In the large-cluster regime (small $f_a$ and large $\kappa_b$), the change is less dramatic as collectively moving clusters are seen both with and without reversals. However, the size of the clusters is considerably decreased by the reversals.

\begin{figure}[t!]
    \centering
   \includegraphics[width=\linewidth]{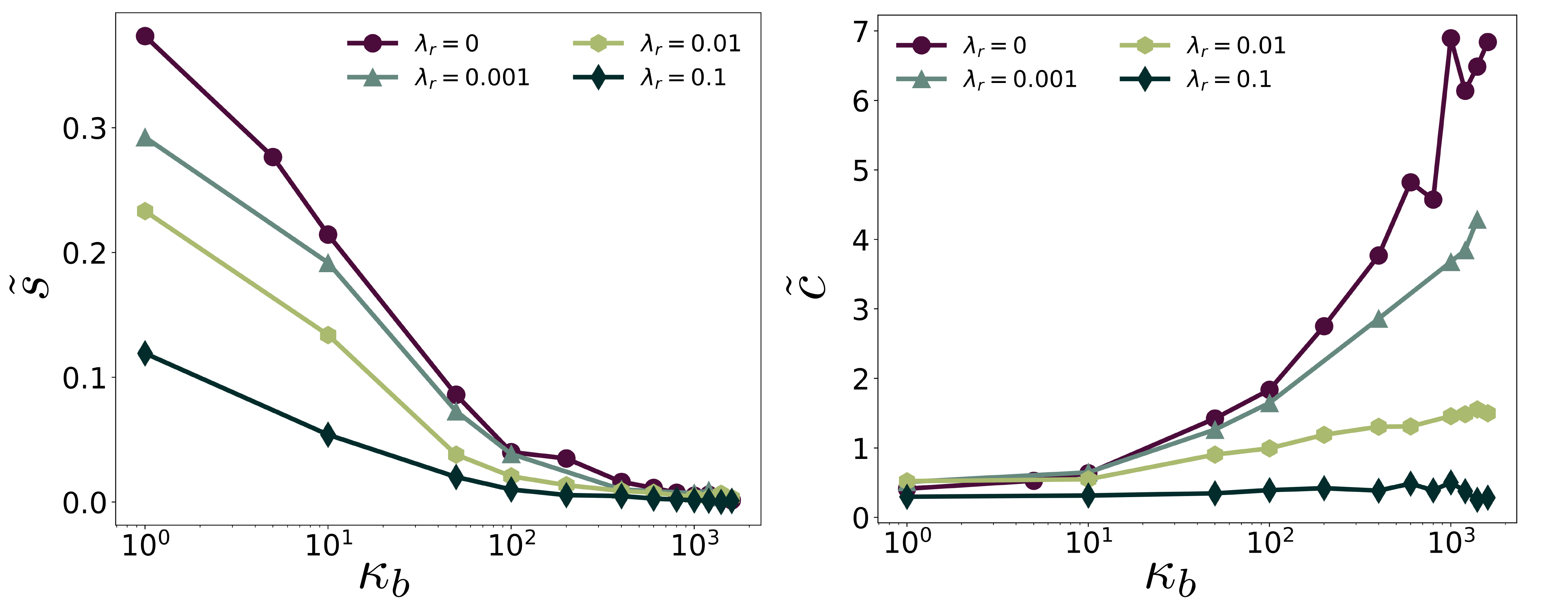}
    \caption{Spiral and cluster ratio for different reversal rates $\lambda_r$ at an active force, $f_a=50$.  }
    \label{fig:reversal-sc}
\end{figure}
%%%%%%%%%%%%%%%%%%%%%%%%%%%%%%%%%%%%%%%%%%%%%%%%%%%%%%%%%%%%%%%%%%%%%%
%                                                                    %
%           Unwinding of spirals and declustering                    %
%                                                                    %    
%%%%%%%%%%%%%%%%%%%%%%%%%%%%%%%%%%%%%%%%%%%%%%%%%%%%%%%%%%%%%%%%%%%%%%

\subsubsection{Unwinding and declustering}
%%%%%%%%%%%%%%%%%%%%%%%%%%%%%%%%%%%%%%%%%%%%%%%%%%%%%%%%%%%%%%%%%%%%%%%%%%%%%%%%%%%%%%%%%%%%%%%%%%%%%%%%%%%%%%%%%%%%%%%%%%%
\begin{figure*}[ht!]
    \centering
   \includegraphics[width=0.8\linewidth]{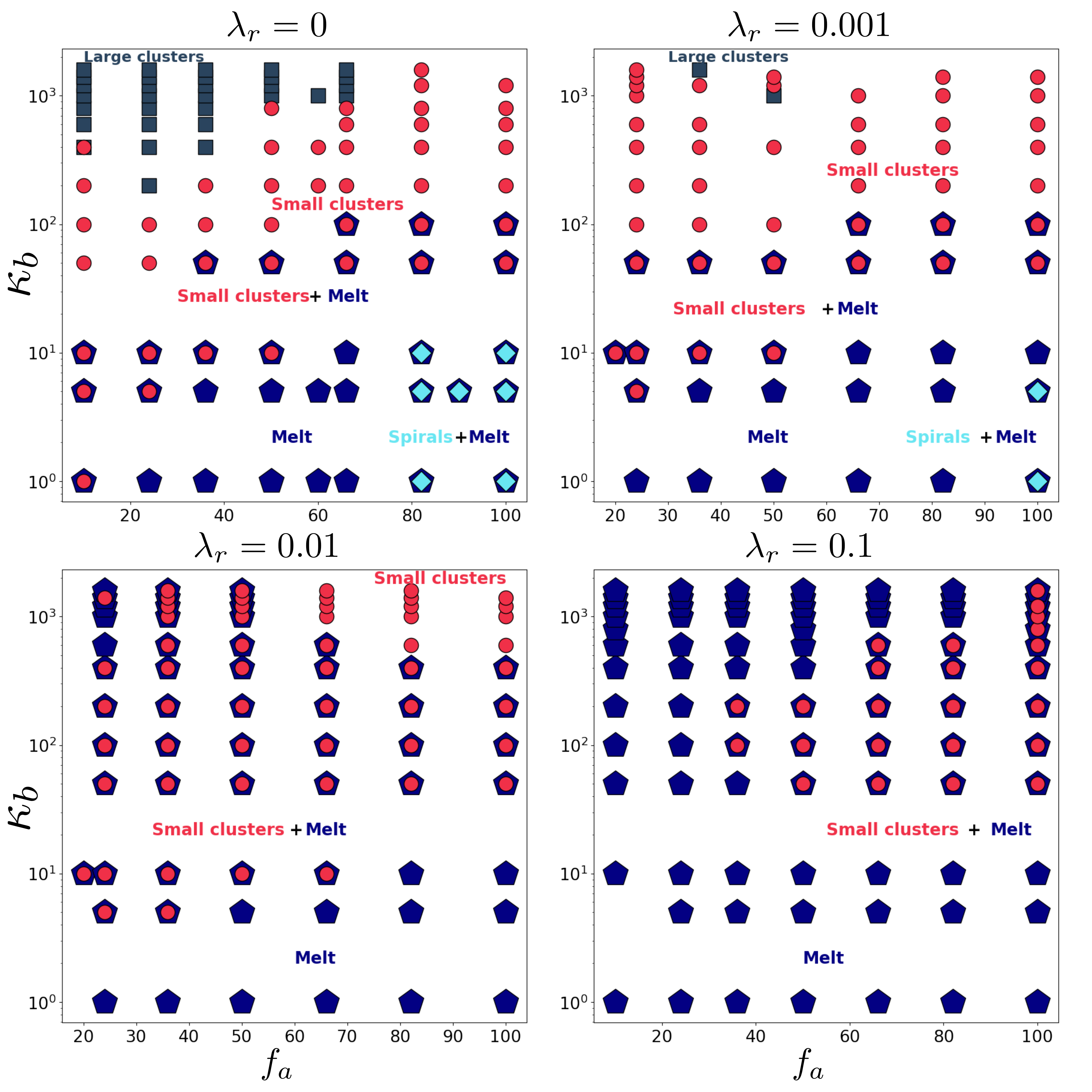}
    \caption{\textbf{Diagram of state for reversing  self-propelled filaments.} The state diagram is constructed as a function of the self-propulsion force $f_a$ and the bending stiffness $\kappa_b$ and shown for four different values of the reversal rate $\lambda_r$. States are marked by different symbols and different colors as in Figure 6.
    When the reversal rate increases, on the one hand the tendency of formation of large clusters decreases, on the other hand the propensity of spiral formation in the system is progressively suppressed.} 
    \label{fig:phase-diagram-reverse}
\end{figure*}
%%%%%%%%%%%%%%%%%%%%%%%%%%%%%%%%%%%%%%%%%%%%%%%%%%%%%%%%%%%%%%%%%%%%%%%%%%%%%%%%%%%%%%%%%%%%%%%%%%%%%%%%%%%%%%%%%%%%%%%%%%%
%%%%%%%%%%%%%%%%%%%%%%%%%%%%%%%%%%%%%%%%%%%%%%%%
\begin{figure*}[ht!]
    \centering
   \includegraphics[width=\linewidth]{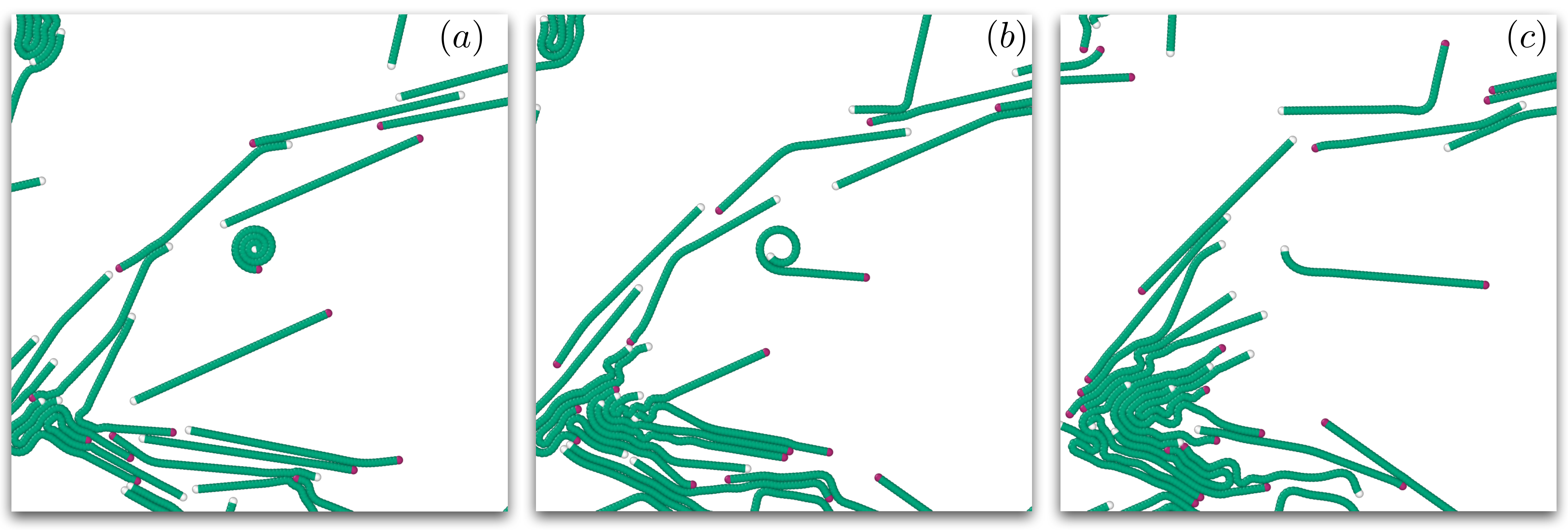}
    \caption{\textbf{Unwinding of a spiral due to the reversal dynamics:} Panel (a) shows a filament in the spiral state in the center. In Panel (b) a reversal event takes place. Panel (c): The spiral unwinds and the filament resumes the straight configuration. The white monomer is the head of the filament, the green is the body and the purple is the tail. } 
    \label{fig:spiral-unwind}
\end{figure*}
%%%%%%%%%%%%%%%%%%%%%%%%%%%%%%%%%%%%%%%%%%%%%%%%%%%%%%%%%%%%%%%%%%%%%%%%%%%%%%%%%

To quantify the weakening of spiral and cluster formation, we determined again the spiral and cluster ratio, for a large range of reversal rates $\lambda_r$, for an intermediate value of $f_a=50$. In Fig.~\ref{fig:reversal-sc}, we plot the two quantities as functions of the bending stiffness, comparing different reversal rates. These results confirm the qualitative observation from the snapshots: Reversals reduce both spiral and cluster abundance, for all values of the bending stiffnesses. Higher reversal rates result in stronger reduction.

Having noticed this stark effect of the reversal rate on pattern formation by active filaments, we next investigate the effects of the reversal rate $\lambda_r$ on the state diagram in the $\kappa_b$-$f_a$ space. The results are shown in Fig.~\ref{fig:phase-diagram-reverse}, for three non-zero values of $\lambda_r$, in comparison to the case without reversals. Already for the smallest non-zero reversal rate that we considered, $\lambda_r=0.001$, the spiral state as well as the large-cluster state are strongly diminished. Increasing the reversal rate further, also the small-cluster regime moves towards larger bending stiffnesses and larger self-propulsion forces. 

%%%%%%%%%%%%%%%%%%%%%%%%%%%%%%%%%%%%%%%%%%
The suppression of spirals by reversals is rather straightforward. As mentioned earlier, in the absence of reversals, spirals dissolve only through collision with other filaments. Reversals add a second mechanism by the change in the polarity (head-tail direction): When the reversal occurs, self-propulsion is not directed inward, but rather outward, and thus the spiral unwinds (see Fig.~\ref{fig:spiral-unwind}) unless another reversal occurs before unwinding was complete. For small reversal rates $\lambda_r < f_a/\zeta L$, where a filament propagates by more than its body length between reversals, this is expected to happen. However, unwinding is faster than winding, because of the additional release of bending energy, such that also repeated reversals lead to a net unwinding.

Likewise, reversals add an additional mechanism for filaments to leave a cluster. Clustered filaments move collectively in one direction, so that any filament that reverses is likely to move out of the cluster. This shifts the balance between filaments joining a cluster and filaments leaving a cluster, resulting in smaller clusters. For very large reversal rates, clusters disappear entirely and the system becomes isotropic again. In this limit, the filaments do not show active directed motion any more, but rather move back and forth in a fashion similar to passive diffusion. Thus a melt state like for passive polymers is indeed expected, but the disorder results from an entirely non-equilibrium mechanism.

%%%%%%%%%%%%%%%%%%%%%%%%%%%%%%%%%%%%%%%%%%%%%%%%%%%%%%%%%%%%%%%%%%%%%%
%                                                                    %
%                                Conclusion                          %
%                                                                    %    
%%%%%%%%%%%%%%%%%%%%%%%%%%%%%%%%%%%%%%%%%%%%%%%%%%%%%%%%%%%%%%%%%%%%%%

\section{Summary and Conclusion}

In this study, we used particle-based simulations of self-propelled filaments to study their pattern formation and, specifically, the impact of direction reversal on these patterns. The simulations mimic bacterial gliding motility on surfaces and their assembly into clusters and swarms. In contrast to the well-studied spherical or rod-like self-propelled particles, we considered long and flexible filaments.

First, we discussed the patterns that could form in the absence of direction reversals,  modulating the bending stiffness ($\kappa_b$) and self-propulsion force ($f_a$) of the filaments at a relatively low filament density. The results, in agreement with earlier work \cite{Prathyusha2018,duman2018}, reflect a competition between the self-interaction of filaments, possible due to their flexibility, and the interaction between filaments. At low bending stiffness and high self-propulsion force, the self-interactions are dominant, leading to the formation of spirals. However, when the bending stiffness increases, the self-interaction is reduced and steric interactions between filaments become more important. In that case, filaments form clusters through collisions and move collectively in these clusters. Based on systematic simulations, we have determined a diagram of states in the $\kappa_b$-$f_a$ space.

Then, to understand the impact of stochastic direction reversals, we added a reversal rate to the filaments' dynamics,  mimicking the run-reverse movement of bacteria, and examined the effects of this additional control parameter on the formation of motility patterns. In general, reversals counteract both clustering and spiral formation, thus resulting in more isotropic structures. Both negative effects can be understood via dynamical pathways these structures: First, reversals add a second mechanism to interrupt the spooling of spirals and induce unspooling. Second, it also adds new mechanisms for filaments to leave a cluster, as filaments typically leave a coherently moving cluster upon reversal. The additional mechanism for leaving clusters results in smaller clusters compared to the no-reversal reference case. Finally, high reversal rates result in purely back-and-forth motion, similar to a passive scenario.

Our results show that the direction reversals have a great impact on the patterns of microorganisms in the casting media or other systems. The direction reversals have unpleasant effects (like the destruction of spirals and clustering) on the patterns formed by the bacteria, which could also affect the dynamics of microorganisms in the system. Since the direction reversal has a very unique effect on the self-assembly of active filaments, it can also be extended to a higher density to study the effects of the reversal rate on the pattern formed in that regime with/without obstacles in the system.

%%%%%%%%%%%%%%%%%%%%%%%%%%%%%%%%%%%%%%%%%%%%%%%%%%%%%%%%%%%%%%%%%%%%%%
%                                                                    %
%                             Acknowledgement                        %
%                                                                    %    
%%%%%%%%%%%%%%%%%%%%%%%%%%%%%%%%%%%%%%%%%%%%%%%%%%%%%%%%%%%%%%%%%%%%%%

\section*{Acknowledgement}
We are grateful to Rituparno Mandal for helpful discussions.
This research was conducted within the Max Planck School Matter to Life, supported by the German Federal Ministry of Education and Research (BMBF) in collaboration with the Max Planck Society.

\bibliography{reference}

\end{document}